\documentclass[a4paper,12pt]{article}

\usepackage[hmargin=.7in,vmargin=1.1in]{geometry}
\usepackage{indentfirst}
\usepackage{amsfonts}
\usepackage{mathrsfs}
\usepackage{amsmath}
\usepackage{amssymb}
\usepackage{authblk}
\usepackage{cite}
\usepackage{xcolor}
\usepackage{mathtools}
\usepackage{tensor}
\usepackage{physics}
\usepackage{graphicx}
\usepackage{bm}
\usepackage{upgreek}
\usepackage{braket}
\usepackage{color,soul}
\usepackage{csquotes}
\usepackage{caption}
\usepackage{subcaption}
\usepackage{tabularx}
\usepackage[section]{placeins}
\usepackage{enumerate}
\usepackage{comment}

\newcommand\mi{\mathrm{i}}

\newcommand{\dif}{\rd }

\newcommand{\rd}{\mathrm{d}}
\newcommand{\re}{\mathrm{e}}

\usepackage[bookmarksnumbered=true,bookmarksopen=true]{hyperref}
 \hypersetup{colorlinks,
             linkcolor=[rgb]{0,0.3,0.6}, 
             citecolor=[rgb]{0,0.3,0.6}, 
             urlcolor=[rgb]{0,0.3,0.6}}

\title{\Large Thermodynamics for regular black holes as 
intermediate thermodynamic states and quasinormal frequencies 
}

\author[1]{Bai-Hao Huang\thanks{huangbaihao@mail.nankai.edu.cn}}
\author[1]{Han-Wen Hu\thanks{huhanwen@mail.nankai.edu.cn}}
\author[1]{Liu Zhao \thanks{Corresponding author, lzhao@nankai.edu.cn,}}

\affil[1]{\normalsize{\em School of Physics, Nankai University, 
Tianjin 300071, China}}

\numberwithin{equation}{section}

\date{}

\begin{document}

\maketitle

\begin{abstract}

The thermodynamics for regular black holes (RBHs) is considered 
under the restricted phase space (RPS) formalism. 
It is shown that the RPS formalism seems to hold for RBHs, 
however, in order for the extensive thermodynamic parameters to be 
independent from each other, the RBHs need to be viewed as intermediate 
thermodynamic states in a larger class of black holes (BHs) which
admit both regular and singular states. This idea is checked for 
several classes of BHs. In particular, for the electrically 
charged Hayward class BHs, it is shown that the regular states 
can either be thermodynamically stable or unstable, depending on 
the amount of charges carried by the BHs. The quasinormal 
frequencies for the Hayward class BHs are also analyzed, 
and it turns out that, even for the thermodynamically unstable
regular states, the dynamic stability still holds, at least under
massless scalar perturbations.

\vspace{1em}


\end{abstract}


\section{Introduction}
\label{sec:intr}
Black holes (BHs) are important objects predicted by general relativity and 
their existence is verified by gravitational wave observations. Theoretically, 
under the strong energy condition and the weak cosmic censorship conjecture,  
singularities are inevitable inside the event horizons of the BHs. 
However, the existence of singularities signify the incompleteness of spacetime 
and the failure of all physics laws \cite{Markov:1982ld,Wald:1984rg}, 
there have been long-lasting efforts in studying regular black holes (RBHs)
since the 1970s \cite{Bardeenblack}. Besides baldly replacing the singular core of the 
black hole metric with a regular one, there are also some theoretical reasons 
for envisaging the replacement of singular BHs by RBHs. To name a few of the 
underlying reasonings, the ultraviolet incompleteness of general relativity 
calls for a smallest spacetime resolution of the Planck size, and 
also the strong energy condition may be weakened 
in certain circumstances. Nowadays, a significant amount of interest has been paid toward
a particular class of RBHs which arise as solutions to the field equations 
of general relativity coupled with nonlinear electrodynamics (NED) 
\cite{Ayon-Beato:2000mjt,Ayon-Beato:1998hmi,Ayon-Beato:2004ywd,Hayward:2005gi}.

Besides pure theoretical motivations, the recent development in observational techniques,
such as the Event Horizon Telescope \cite{EventHorizonTelescope:2019dse,
EventHorizonTelescope:2019uob}, the LISA project\cite{Barausse:2020rsu} and the 
LIGO-VIRGO-KAGRA coillaborations \cite{LIGOScientific:2016sjg,LIGOScientific:2016vlm}, 
also provides the possibility to verify the validity of various modified theories 
of gravity. Therefore, it is meaningful to calculate the shadow and the 
quasinormal frequencies of RBHs to assist in the analysis of experimental data. 
The study of RBHs may also provide a potential solution for the information loss 
paradox, which seems to be related to the singularity 
in the core of BHs \cite{Murk:2023vdw}.

In spite of various progresses in the study of RBHs, 
the thermodynamic description for RBHs remains confusing. 
Under the guidance of Bardeen and Hawking's work \cite{Bardeen:1973gs}, 
Rasheed redefined the electric charge, magnetic charge, electric potential, 
and magnetic potential, and gave the first law of RBHs\cite{NonLinear1997}. 
However, it fails to hold in the Bardeen and Hayward RBHs
\cite{Corrected2014,Zhang:2016ilt}. Ref.\cite{Corrected2014} suggested that 
$\rd M$ in the first law needs a coefficient that is not always 
equal to 1, which is followed up by many authors \cite{Corrected2014,Zhang:2016ilt,
Lan:2021ngq,Fan:2016hvf,Fan:2016rih} and leads to a very different form of the 
first law and Smarr formula. Such modifications suffer the problems of 
parameter non-independence \cite{Lan:2023cvz,depaula2023electrically,li2023regular,Hu:2023iuw} and non-extensiveness.

In this work, we employ the restricted phase space (RPS) formalism
\cite{Zeyuan:2021uol,Gao:2021xtt} 
to analyze the thermodynamic behaviors of RBHs. We find that the first law 
and the Smarr formula of RBHs are very similar to those of singular BHs, 
however the regularity condition calls for a constraint between the mass and the charge, 
which also results in a parameter non-independence problem. 
This situation leads us to suggest that RBHs may not be thermodynamically 
self-consistent. In order to understand the thermodynamic behavior of RBHs, 
one might need to view them as some intermediate states in the thermodynamic 
processes of a larger class of BHs involving both singular and regular states. 
One of the major motivation for this work is to 
verify this idea. Meanwhile, we are also interested in the 
stability of RBHs. We will study the stability of RBHs from both 
thermodynamic and dynamic perspectives. The thermodynamic stability 
is analyzed by considering the behaviors under the RPS formalism, while the dynamical
stability will be analyzed by studying the quasinormal frequencies.

This paper is organized as follows. In Sec.~\ref{se:Outline}, some 
issues in existing attempts for thermodynamics
of RBHs are outlined. In Sec.~\ref{se:RPST}, 
we derive the first law and Smarr formula under the RPS formalism for four 
different classes of black hole solutions which can become regular at 
some specific choices of parameters, i.e. the Bardeen class, the Hayward class, 
the Bardeen-AdS class \cite{Fan:2016hvf}, and the new class of RBHs presented 
recently in ref.\ \cite{li2023regular}. The isocharge $T-S$ processes for
the electrically charged Hayward class BHs are analyzed 
in Sec.~\ref{se:Thermodynamic} under the RPS formalism, which indicates that
there can be at most a single thermodynamically stable branch 
which exists only in a limited range of temperature and only for BHs carrying 
some supercritical values of electric charge. We also study the thermodynamic stability 
of the regular states in the wider class of charged Hayward BHs in 
Sec.~\ref{se:Thought}, using the idea of considering RBHs as 
intermediate states. Sec.~\ref{sec:qnf} is devoted to the study of quasinormal 
frequencies (QNFs) for the whole class of charged Hayward BHs, with emphasis 
on the dynamic stability of the regular states.
Finally, we summarize the results in Sec.~\ref{se:Conclusion}. 
Throughout this paper, we use the subscript ``reg'' to denote the thermodynamic 
quantities at the regular states.

\section{Outline of issues in existing attempts for thermodynamics
of RBHs}\label{se:Outline}

We begin by presenting the action of Einstein gravity coupled with NED 
\cite{Pellicer:1969cf,Bronnikov:2000vy},
\begin{equation}\label{eq:action}
S=\frac{1}{16\pi G}\int_{\mathcal{M}}
{\rd ^4x\sqrt{-g}\left[ R-\mathcal{L} \left( F \right) \right]},
\end{equation}
where $F\equiv F_{\mu \nu }F^{\mu \nu}$ is electromagnetic invariant
and $\mathcal{L} \left( F \right)$ represents the Lagrangian density of the NED. 
There can be different choices for $\mathcal{L} \left( F \right)$ 
which lead to different classes of RBHs. We will specify the concrete form
of $\mathcal{L} \left( F \right)$ when it becomes necessary.

It is worth mentioning that sometimes the NED model is specified by the 
Hamiltonian-like density \cite{Fan:2016hvf,Bronnikov:2017tnz}
\begin{align}
\mathcal{H}(P)=2F\mathcal{L}_F-\mathcal{L}(F)
\end{align}
instead of the Lagrangian density $\mathcal{L}(F)$, where 
\begin{align}
P_{\mu\nu}=\mathcal{L}_FF_{\mu\nu},\quad
\mathcal{L} _F=\partial \mathcal{L}/\partial F, \quad
P=P_{\mu\nu}P^{\mu\nu}.
\end{align} 
It can be shown that 
\begin{equation}
\mathcal{H}_P=\frac{\dif\mathcal{H}}{\dif P}=\frac{1}{\mathcal{L}_F}.
\end{equation}

The equations of motion which arise as the stationary conditions for 
the action \eqref{eq:action} read
\begin{equation}
G_{\mu\nu}=R_{\mu\nu}-\frac{1}{2}g_{\mu\nu}R=8\pi G T_{\mu\nu},\quad 
\nabla_\mu P^{\mu\nu}=0,
\end{equation}
where 
\begin{equation}
T_{\mu\nu}=\frac{1}{4\pi G}\left(\mathcal{L}_F F_{\mu \rho} {F_{\nu}}^{\rho}
-\frac{1}{4}g_{\mu\nu}\mathcal{L}\right).
\end{equation}
We will be interested in the static, spherically symmetric metrics of the form 
\begin{equation}
\rd s^2=-f(r)\rd t^2+\frac{\rd r^2}{f(r)}
+r^2(\rd \theta^2+\sin^2\theta\ \rd \varphi^2).
\end{equation}
Black hole solutions of the above form can carry either electric or 
magnetic charges. The field strengths obeying the Maxwell-like equation 
$\nabla_{\mu}P^{\mu\nu}=0$ 
and the Bianchi identity ${\nabla _{\mu}}^*F^{\mu \nu}=0$ in these two 
cases take the form 
\begin{equation}
P_{tr}=-\frac{Q_e}{r^2},\quad F_{\theta\varphi}=Q_m\sin\theta,
\end{equation}
where $Q_e$ and $Q_m$ are respectively the electric and magnetic charges 
defined via \cite{NonLinear1997}
\begin{equation}\label{eq:Q_m-Q_e}
Q_m=\frac{1}{4\pi }\int_{\partial \Sigma}{F_{ab}},\quad 
Q_e=\frac{1}{4\pi }\int_{\partial \Sigma}{^*P_{ab}},
\end{equation}
in which $\partial \Sigma$ is the boundary of the spacelike hypersurface $\Sigma$ 
with the timelike Killing vector field $\xi^a$ acting as its normal vector. 
As usual, one can introduce the electric and magnetic fields
\[
E_a =\xi^b F_{ab},\quad  H_a= - \xi^b \, {}^\ast P_{ab},
\]
together with the corresponding potentials,
\begin{align}
E_a&=-\nabla_a \Phi,\quad H_a=-\nabla_a \Psi.
\label{eq:Phi-Psi}
\end{align}

Considering that Rasheed's first law and Smarr formula \cite{NonLinear1997} do not 
work for the RHBs, Zhang and Gao suggested that each parameter involved in the Lagrangian of 
the matter field may introduce an extra term in the first law of RBHs 
\cite{Zhang:2016ilt}. For instance, the first law and Smarr formula for the 
magnetically supported Bardeen BHs are 
modified as
\begin{align}
\rd  M &=\frac{\kappa}{8\pi }\rd  A
+\Psi _{\rm H}\rd  Q_m+K_{q} \rd  Q_m+K_{M}\rd  M,  
\label{2.8}
\\
M&=\frac{\kappa}{4\pi } A+\Psi _{\rm H} Q_m+K_{q}  Q_m+K_{M} M,
\end{align}
where $\Psi_{\rm H}$ is the magnetic potential on the event horizon, 
and $K_{q},~K_{M}$ are defined as
\begin{equation}
K_{q} =\frac{1}{4}\int_{r_{\rm H}}^{\infty}
{\rd r\sqrt{-g}\frac{\partial \mathcal{L}}{\partial Q_m}}, \quad 
K_{M} =\frac{1}{4}\int_{r_{\rm H}}^{\infty}{\rd r\sqrt{-g}
\frac{\partial \mathcal{L}}{\partial M}},
\end{equation}
and $\kappa$ and $A$ are respectively the surface gravity and the area of the event horizon. 
Since the model under investigation is general relativity coupled to NED, 
the Bekenstein-Hawking entropy formula is valid and the term $\displaystyle 
\frac{\kappa}{8\pi }\rd  A$ \vspace{3pt} 
in eq.\eqref{2.8} can be replaced by $T\rd S$ as usual.
The problem of the above proposal lies in that the meaning of $K_{q},~K_{M}$ are unclear
and that the term $\rd M$ acquires an extra coefficient $(1-K_M)$ in the first law.

Alternatively, Wang and Fan \cite{Fan:2016hvf,Fan:2016rih} suggested a different 
proposal which brings the model parameter $\alpha$ that appears in the Lagrangian 
density of the NED into the first law and the Smarr formula. For a class of RBHs 
carrying both electric and magnetic charges, the resulting first law and the Smarr formula
read
\begin{align}\label{eq:Wang-first-law}
\rd  M_{\rm ADM}&=\frac{\kappa}{8\pi }\rd  A+\Phi _{\rm H}\rd  Q_e
+\Psi _{\rm H}\rd  Q_m+\Pi \rd  \alpha
\\
M_{\rm ADM}&=\frac{\kappa}{4\pi }\rd  A+\Phi _{\rm H} Q_e
+\Psi _{\rm H} Q_m+2\Pi \alpha,
\end{align}
where $\Psi_{\rm H}$ is as before, $\Phi_{\rm H}$ is the electric potential on the event horizon, 
and $\Pi$ is defined as
\begin{equation}
\Pi =\frac{1}{4}\int_{r_{\rm H}}^{\infty}{\rd r\sqrt{-g}
\frac{\partial \mathcal{L}}{\partial \alpha}}.
\end{equation}
The inclusion of a variable negative cosmological constant may lead to an extra  
$V\rd P$ term \cite{Fan:2016hvf,Fan:2016rih}.
Similar works appear in Ref.\cite{Bokulic:2021dtz,Gulin:2017ycu}, 
where generalizations of the first law and Smarr formula for BHs 
with NED are respectively deduced from the thought that each 
parameter $\alpha_i$ in Lagrangian density of the NED introduces 
an extra term in the first law and Smarr formula of RBHs
and keep these parameters constant within fixed spacetime (i.e.$\nabla_a\alpha_i=0$).
The problem lies in that model parameter $\alpha$ is related to the magnetic charge 
$Q_m$ and the ADM mass $M_{\rm ADM}$ via \cite{Lan:2023cvz}
\begin{equation}\label{eq:alpha-demand}
\alpha=\frac{8G^2Q^6_m}{M_{\rm ADM}^4},
\end{equation}
therefore, there is a problem of parameter non-independence. 
Moreover, changing $\alpha$ implies changing the underlying model, and hence there is 
also an ensemble of theories problem.

In spite of the numerous attempts mentioned above, it appears that a consistent 
thermodynamic description for RBHs is still missing. In the next section, we will try to
establish a self consistent thermodynamic description for RBHs using the recently
proposed RPS formalism. It will be seen that, in order to have a self consistent 
thermodynamic description, it is better to view the RBHs as some intermediate states
for a larger class of BHs which are singular at wider range of values 
of the parameters, whilst become regular at some specific values of parameters. 
In this way, all issues suffered by the previous proposals are avoided.

\section{RPS formalism for RBHs}\label{se:RPST}

In this section, we will employ the RPS formalism for describing the thermodynamics of RBHs. 
The RPS formalism is inspired by Visser's holographic thermodynamics \cite{Visser:2021eqk}, 
but with an important difference, i.e. the cosmological constant must be kept
invariant. This allows for the RPS formalism to be applicable to much wider classes of 
black hole solutions without urging an AdS asymptotics \cite{Wang:2021cmz,Zhao:2022dgc}. 
Moreover, the effective number of microscopic degrees of freedom, $N=L^{n-2}/G$, 
and the corresponding conjugate chemical potential, $\mu=GTI_{\rm E}/L^{n-2}$, are 
both defined independently without using the Euler relation. 
All these features suggest that the RPS formalism may also be applicable
to RBHs, and it will be clear that it is indeed the case.

\subsection{RPS formalism for the electrically charged Bardeen class BHs}\label{subse:Bardeen}

First let us consider the RPS formalism for the electrically charged 
Bardeen class BHs. 

The Bardeen class BHs are supported by an NED with the Hamiltonian-like density
\begin{align}\label{eq:Bardeen-Hamiltonian-like density}
\mathcal{H}(P)=-\frac{4s\left( -\alpha P \right)^{5/4}}
{\alpha \left( 1+\sqrt{-\alpha P} \right) ^{1+s/2}},
\end{align}
where $s$ is a dimensionless constant. Under the assumption that the black hole solution
is supported purely by an electric field, Ref.\cite{Toshmatov:2018cks} suggested
that the shape function $f(r)$ in the metric should take the form
\begin{equation}\label{eq:f(r)-Bardeen}
f(r)=1-\frac{2Gm(r)}{r},\quad m(r)
=M_g-\frac{q^3}{G\alpha}\left[1-\frac{r^s}{\left(r^2+q^2\right)^{s/2}} \right],
\end{equation}
where the integration constant $q$ is related to the electric charge $Q_e$ via
\begin{align}
Q_e&=\frac{q^2}{G\sqrt{2\alpha}}.
\label{eq3.6}
\end{align} 
The form \eqref{eq:f(r)-Bardeen} of the shape function ensures that 
the ADM mass is always equal to the integration constant $M_g$, 
i.e. $M_{\rm ADM}=M_g$. 

The Kretschmann invariant for the above solution reads
\begin{align}
R^{\mu \nu \rho \kappa}R_{\mu \nu \rho \kappa}
&=\frac{2}{r^6}\left[ 2\left( GM_g-\frac{q^3}{\alpha} \right) 
+2\frac{q^3r^s}{\alpha}\frac{\left[ r^2-\left( s-1 \right) q^2 \right]}
{\left( r^2+q^2 \right) ^{s/2+1}} \right] ^2\nonumber
\\
&+\frac{4}{r^6}\left[ 2\left( GM_g-\frac{q^3}{\alpha} \right)
+\frac{2r^s}{\alpha}\frac{q^3}{\left( r^2+q^2 \right) ^{s/2}} \right] ^2\nonumber
\\
&+\frac{1}{r^6}\left[ 4\left( GM_g-\frac{q^3}{\alpha} \right)
+\frac{2q^3r^s}{\alpha}\frac{\left[ 2r^4-\left( 5s-4 \right) r^2q^2
+\left( s-1 \right) \left( s-2 \right) q^4 \right]}
{\left( r^2+q^2 \right) ^{s/2+2}} \right] ^2.
\end{align}
It is obvious that the solution has an intrinsic singularity at $r=0$ 
when $GM_g\ne q^3/\alpha$, and it is regular at $r=0$ if 
$GM_g=q^3/\alpha,~s\geqslant 3$. When $s=3$, this regular solution 
corresponds to the original Bardeen RBH \cite{Bardeenblack}. 
For this reason, the whole class of black hole solution specified by 
the shape function \eqref{eq:f(r)-Bardeen} will be referred to as Bardeen class BHs. 

For generic choice of parameters (i.e. without imposing the regularity condition 
$GM_g=q^3/\alpha,~s\geqslant 3$), the temperature $T$ and entropy $S$ are given by
\begin{equation}\label{eq:Bardeen-T}
T=\frac{1}{4\pi }\left( \frac{\partial f}{\partial r} \right) _{r=r_{\rm H}}
=\frac{1}{4\pi r_{\rm H}}-\frac{s q^5 r_{\rm H}^{s-2}}
{2\pi \alpha \left( r_{\rm H}^{2}+q^2 \right) ^{s/2+1}},\quad 
S=\frac{\pi r^2_{\rm H}}{G},
\end{equation}
where $r_{\rm H}$ is the radius of the event horizon which arises as the largest root 
of $f(r)$. The value of the electric potential $\Phi_{\rm H}$ 
on the event horizon reads
\begin{align}
\Phi_{\rm H}&=\frac{q}{\sqrt{2\alpha}}
\left[ 3-\frac{3r_{\rm H}^{s}}{\left( r_{\rm H}^{2}+q^2 \right) ^{s/2}}
+\frac{sq^2r_{\rm H}^{s}}{\left( r_{\rm H}^{2}+q^2 \right) ^{1+s/2}} \right].
\end{align}
The on-shell Euclidean action $I_{\rm E}$ corresponding to the above solution is 
\cite{York:1986it,Gibbons:1976ue,Lan:2023cvz}
\begin{align}
I_{\rm E}&=-\frac{1}{16\pi G}\int_{\mathcal{M}}{\sqrt{g}
\left[ R-\mathcal{L}(F) \right]\rd ^4x}
-\frac{1}{8\pi G}\int_{\partial \mathcal{M}}{\sqrt{h}\left( K \right)\rd ^3x}
+\frac{1}{8\pi G}\int_{\partial \mathcal{M}}{\sqrt{h}\left( K_0 \right)\rd ^3x}\nonumber
\\
&=\frac{M_g\beta}{2}-\frac{\beta q^3}{G\alpha}
+\frac{\beta q^3}{G\alpha}\frac{r_{\rm H}^{s}}{\left( r_{\rm H}^{2}+q^2 \right) ^{s/2}}.
\end{align}
Therefore, according to the general formula from the RPS formalism, we can introduce 
the chemical potential $\mu$ and the effective number of microscopic degrees of 
freedom $N$ as follows,
\begin{equation}\label{eq:Bardeen-mu}
\mu=\frac{TI_{\rm E}}{N},\quad N=\frac{L^2}{G},
\end{equation}
where $L$ is a constant length scale.

By straightforward calculation, the following first law and Euler relation 
can be verified to hold, 
\begin{align}
\rd M_{\rm ADM}&=T\rd S+\Phi_{\rm H}\rd Q_e+\mu \rd N, \label{dM1}
\\
M_{\rm ADM}&=TS+\Phi_{\rm H} Q_e+\mu N. \label{M1}
\end{align}

Now let us restrict ourselves to the regular case with $M_g=q^3/G\alpha$. 
We can describe the regular value 
$M_{\rm reg}$ of $M_{\rm ADM}$ by use of the equation $f(r_{\rm H})=0$, 
\begin{equation}\label{eq:ADM-Q_e}
M_{\rm reg}=\frac{q^3}{G\alpha}
=\frac{\left( r_{\rm H}^{2}+q^2 \right) ^{s/2}}{2Gr_{\rm H}^{s-1}}.
\end{equation}
Here, $\alpha$ is required to remain as a constant, 
\begin{equation}\label{eq:alpha-command1-Bardeen}
\alpha =\frac{2q^3r_{\rm H}^{s-1}}{\left( r_{\rm H}^{2}+q^2 \right) ^{s/2}}
=\mathrm{const}.
\end{equation}
Consequently we have
\begin{equation}\label{eq:alpha-command-Bardeen}
\rd r_{\rm H}=\frac{r_{\rm H}}{q}\frac{3r_{\rm H}^{2}
-\left( s-3 \right) q^2}{r_{\rm H}^{2}-\left( s-1 \right) q^2}\rd q.
\end{equation}

Substituting the condition $M_g=q^3/G\alpha$ into eqs.\eqref{eq:Bardeen-T} 
and \eqref{eq:Bardeen-mu}, the temperature and chemical potential in the 
regular case can be rewritten as 
\begin{align}
T_{\rm reg}=\frac{q^3r_{\rm H}^{s-2}\left[ r_{\rm H}^{2}
-\left( s-1 \right) q^2 \right]}{2\pi \alpha \left( r_{\rm H}^{2}+q^2 \right)^{s/2+1}},
\quad 
\mu_{\rm reg}=\frac{G}{L^2}\left[ -\frac{q^3}{2G\alpha}
+\frac{q^3}{G\alpha}\frac{r_{\rm H}^s}{\left( r_{\rm H}^2+q^2 \right) ^{s/2}} \right].
\end{align}
By straightforward calculation, it can be verified that following first law and 
Euler relation hold in the regular case,
\begin{align}
\rd M_{\rm reg}&=T_{\rm reg}\rd S+\Phi_{\rm H}\rd Q_{e}
+\mu_{\rm reg} \rd N,\label{eq:RPST-first-law}
\\
M_{\rm reg}&=T_{\rm reg}S+\Phi_{\rm H} Q_{e}
+\mu_{\rm reg} N,
\label{eq:RPST-Euler-relation}
\end{align}
where the differential relation \eqref{eq:alpha-command-Bardeen} needs to be employed
while verifying the above relations. 

Although eqs.~\eqref{eq:RPST-first-law} and 
\eqref{eq:RPST-Euler-relation} are very similar in form to 
eqs.~\eqref{dM1} and \eqref{M1}, their meanings are completely different. 
Eqs.~\eqref{dM1} and \eqref{M1} describe the extensive thermodynamics
for the whole class of electrically charged Bardeen class BHs, while 
eqs.~\eqref{eq:RPST-first-law} and \eqref{eq:RPST-Euler-relation} 
describe only the thermodynamics of the regular Bardeen class BHs. 
Inserting eq.\eqref{eq3.6} into the regularity condition 
\eqref{eq:ADM-Q_e}, we get
\begin{equation}
\alpha=\frac{8G^2Q^6_{e}}{M_{\rm reg}^4},
\end{equation}
which indicates that $Q_{e}$ and $M_{\rm reg}$ are not independent of each other. 
Therefore, eqs.~\eqref{eq:RPST-first-law} and \eqref{eq:RPST-Euler-relation} 
contain exactly the same problem of parameter non-independence as appeared 
in earlier proposals using different formalisms. 
We thus suggest that, instead of 
eqs.~\eqref{eq:RPST-first-law} and \eqref{eq:RPST-Euler-relation}, 
eqs.~\eqref{dM1} and \eqref{M1} should be adopted while describing the thermodynamic
behaviors of RBHs. This in turn suggests that RBHs should be viewed as 
intermediate states for the larger Bardeen class BHs 
specified by the shape function \eqref{eq:f(r)-Bardeen} 
which admit both regular and singular states.

\subsection{RPS formalism for other BHs involving regular states}
\label{subse:Hayward}

In order to support the proposal presented near the end of the last section, 
let us proceed by briefly outlining RPS formalism for other classes of BHs
involving regular states. 

\begin{enumerate}[(i)]
\item Hayward class BHs \cite{Hayward:2005gi}. 
The Hamiltonian-like density of the NED is now given as
\begin{align}
\mathcal{H}(P)=-\frac{4s}{\alpha}
\frac{\left( -\alpha P \right)^{\left( s+3 \right) /4}}
{\left( 1+\left( -\alpha P \right) ^{s/4} \right) ^2},
\end{align}
and the corresponding shape function reads
\begin{align}
f(r)
=1-\frac{2GM_g}{r}-\frac{2q^3}{\alpha r}\left[ \frac{r^s}{ r^s+q^s }-1 \right],
\label{eq:general-f(r)-Hayward}
\end{align}
which is supported by an electric field of charge $Q_e$ and electric potential 
$\Phi_{\rm H}$ on the event horizon,
\begin{align}
Q_e&=\frac{q^2}{G\sqrt{2\alpha}},\label{eq:Hayward-phi-Q}
\\
\Phi_{\rm H}&=\frac{q}{\sqrt{2\alpha}}\left[ 3-\frac{3r_{\rm H}^s}{r_{\rm H}^s+q^s}
+\frac{sr_{\rm H}^sq^s}{\left( r_{\rm H}^s+q^s \right) ^2} \right].
\end{align}

The regularity conditions are still given by $M_g=q^3/G\alpha$ and $s\geqslant 3$,
among which the $s=3$ case is the original Hayward BH. Therefore, the whole class of 
BHs described above will be referred to as Hayward class BHs. 

Besides $Q_e$ and $\Phi_{\rm H}$, the other macroscopic parameters of the 
Hayward class BHs are given as follows,
\begin{align}
T&=\frac{1}{4\pi r_{\rm H}}-\frac{sq^{s+3}r_{\rm H}^{s-2}}
{2\pi \alpha \left( r_{\rm H}^{s}+q^s \right) ^2},\quad 
S=\frac{\pi r^2_{\rm H}}{G},\label{eq:Hayward-T-S}
\\
\mu&=\frac{G}{L^2}\left[ \frac{M_g}{2}-\frac{q^3}{G\alpha}
+\frac{q^3}{G\alpha}\frac{r_{\rm H}^{s}}{r_{\rm H}^{s}+q^s} \right],\quad 
N=\frac{L^2}{G}.\label{eq:Hayward-mu-N}
\end{align}
It can be checked that, for the Hayward class BHs, 
we still have $M_{\rm ADM}=M_g$, and 
the first law \eqref{eq:RPST-first-law} and Euler relation 
\eqref{eq:RPST-Euler-relation} still hold, and the regular states 
should be regarded as intermediate states of the larger Hayward class BHs. 

\item Bardeen-AdS class BHs \cite{Guo:2022yjc,Fan:2016hvf}. 
One can introduce a negative cosmological constant $\Lambda$ into the action for the 
Bardeen class BHs, yielding 
\begin{equation}
S=\frac{1}{16\pi G}\int_{\mathcal{M}}{\rd ^4x\sqrt{-g}
\left[ R-2\Lambda-\mathcal{L} \left( F \right) \right]}. 
\end{equation}
The corresponding electrically supported black hole solutions will be 
referred to as the Bardeen-AdS class BHs, which is specified by the 
shape function 
\begin{equation}\label{eq:f(r)-Bardeen-AdS}
f(r)=1-\frac{2GM_g}{r}-\frac{2q^3}{\alpha r}\left[ \frac{r^s}{\left( r^2+q^2 \right) ^{s/2}}-1 \right] +\frac{r^2}{l^2},
\end{equation}
where $l^2=-3/\Lambda$. 

It follows that the AD mass for this class of BHs is equal to $M_g$. 
Moreover, the regularity conditions remain unchanged, $M_g=q^3/G\alpha,~s\geqslant 3$. 
The macroscopic parameters for this class of BHs are given as follows,
\begin{align}
T&=\frac{1}{4\pi r_{\rm H}}+\frac{3r_{\rm H}}{4\pi l ^2}
-\frac{q^5sr_{\rm H}^{s-2}}{2\pi \alpha \left( r_{\rm H}^{2}+q^2 \right) ^{s/2+1}},
\qquad 
S=\frac{\pi r^2_{\rm H}}{G},
\\
\Phi_{\rm H}&=\frac{q}{\sqrt{2\alpha}}\left[ 3-\frac{3r_{\rm H}^{s}}
{\left( r_{\rm H}^{2}+q^2 \right) ^{s/2}}+\frac{sq^2r_{\rm H}^{s}}
{\left( r_{\rm H}^{2}+q^2 \right) ^{1+s/2}} \right],
\quad 
Q_e=\frac{q^2}{G\sqrt{2\alpha}},
\\
\mu&=\frac{G}{l^2}\left[ \frac{M_g}{2}-\frac{r_{\rm H}^{3}}{2Gl ^2}
-\frac{q^3}{G\alpha}+\frac{q^3}{G\alpha}\frac{r_{\rm H}^{s}}
{\left( r_{\rm H}^{2}+q^2 \right) ^{s/2}} \right],
\quad 
N=\frac{l^2}{G}.
\end{align}
It is easy to verify the first law 
\[
\rd M_{\rm AD}=T\rd S+\Phi_{\rm H}\rd Q_e+\mu \rd N
\]
and Euler relation 
\[
M_{\rm AD}=TS+\Phi_{\rm H} Q_e+\mu N
\]
hold irrespective of whether the regularity conditions are satisfied. 
However, when the regularity conditions hold, the problem of parameter 
non-independence arises once again. Therefore, the regular cases should 
still be regarded as intermediate states of the larger Bardeen-AdS class BHs.

\item Models presented in Ref. \cite{li2023regular}. 
The Hamitonian-like density of the NED field and the simplest shape function 
presented in Ref. \cite{li2023regular} read
\begin{align}
\mathcal{H}(P)=\frac{P}{1-4\alpha P},\quad \left( \alpha \ge 0\right),\quad 
f(r)=1-\frac{2GM_g}{r}+\frac{p^2}{r^2}{_2F_1}
\left[ \frac{1}{4},1;\frac{5}{4};-\frac{2\alpha p^2}{r^4} \right],
\end{align}
where $_2F_1\left[ \frac{1}{4},1;\frac{5}{4};-\frac{2\alpha p^2}{r^4} \right]$ 
denotes a hypergeometric function. The curvature singularity is absent 
when $M_g=\pi p^{3/2}/4G\left(8\alpha\right)^{1/4}$. Besides, the 
ADM mass is equal to the integration constant $M_g$. 
The macroscopic parameters for this class of BHs are given as follows,
\begin{align}
T&=\frac{1}{4\pi r_{\rm H}}-\frac{p^2r_{\rm H}}{4\pi \left( r_{\rm H}^4
+2\alpha p^2 \right)},\quad S=\frac{\pi r_{\rm H}^{2}}{G},
\\
\Phi_{\rm H}&=\frac{3p}{4r_{\rm H}}{_2F_1}\left[ \frac{1}{4},1;\frac{5}{4};
-\frac{2\alpha p^2}{r_{\rm H}^4} \right] 
+\frac{pr_{\rm H}^3}{4\left( r_{\rm H}^4+2\alpha p^2 \right)},\quad 
Q_e=\frac{p}{G},
\\
\mu &=\frac{G}{L^2}\left[ \frac{M_g}{2}
-\frac{p^2}{2Gr_{\rm H}}{_2F_1}\left[ \frac{1}{4},1;\frac{5}{4};
-\frac{2\alpha p^2}{r_{\rm H}^{4}} \right] \right],\quad 
N=\frac{L^2}{G}.
\end{align}
It can be checked by direct calculations that the first law 
\eqref{eq:RPST-first-law} and the Euler relation \eqref{eq:RPST-Euler-relation}
still hold irrespective of whether the regularity condition 
$M_g=\pi p^{3/2}/4G\left(8\alpha\right)^{1/4}$ is obeyed. 
When the regularity condition is fulfilled, parameter 
non-independence problem arises once again, which supports our 
suggestion that the regular BHs should be viewed as intermediate states of 
the larger class of BHs involving both singular and regular states.
\end{enumerate}

\section{Thermodynamic behaviors of Hayward class BHs}
\label{se:Thermodynamic}

Now that the RPS formalism is established for the above mentioned four classes of 
BHs involving regular states, we would like to proceed a little step
by analyzing the thermodynamic behaviors of one out of the four classes. 

We take the Hayward class BHs as example and set $s=3$. 
In order to describe the thermodynamic behaviors, 
it is necessary to rewrite the mass $M(=M_{\rm ADM})$, temperature $T$, 
electric potential $\Phi_{\rm H}$ and chemical potential $\mu$ 
as functions in the additive variables $S,~Q_e,~N$. 
For simplicity of notations, we rescale the electric charge $Q_e$ into 
$Q=Q_e\pi \sqrt{2\alpha}$. Then the parameters $G,~r_{\rm H},~q$ 
can be expressed in terms of $S, Q, N$ by by solving 
eqs.~\eqref{eq:Hayward-T-S},\eqref{eq:Hayward-phi-Q} and \eqref{eq:Hayward-mu-N},
\begin{equation}
G=\frac{L^2}{N},\quad r_{\rm H}=\sqrt{\frac{SL^2}{\pi N}},\quad 
q=\sqrt{\frac{QL^2}{\pi N}}. 
\label{rHq}
\end{equation}
It is then easy to get 
\begin{align}
M(S,Q,N)&=\frac{1}{2L}\sqrt{\frac{NS}{\pi }}
+\frac{N}{L^2\alpha}\left( \frac{QL^2}{\pi N} \right) ^{3/2}
-\frac{N}{L^2\alpha}\left( \frac{QL^2}{\pi N} \right) ^{3/2}
\frac{S^{3/2}}{S^{3/2}+Q^{3/2}},\label{eq:Hayward-M-SQN}
\\
T(S,Q,N)&=\frac{1}{4\pi ^{1/2}}\sqrt{\frac{N}{SL^2}}
-\frac{3Q^3S^{1/2}}{2\pi ^{3/2}\alpha \left( S^{3/2}+Q^{3/2} \right) ^2}
\sqrt{\frac{L^2}{N}},\label{eq:Hayward-T-SQN}
\\
\Phi_{\rm H}(S,Q,N) &=\sqrt{\frac{QL^2}{2\alpha \pi N}}
\left[ \frac{3Q^{3/2}}{S^{3/2}+Q^{3/2}}
+\frac{3S^{3/2}Q^{3/2}}{\left( S^{3/2}+Q^{3/2} \right) ^2} \right],
\label{eq:Hayward-Phi-SQN} 
\\
\mu (S,Q,N)&=\frac{1}{4L}\sqrt{\frac{S}{\pi N}}
-\frac{L}{2\alpha}\left( \frac{Q}{\pi N} \right) ^{3/2}
+\frac{L}{2\alpha}\left( \frac{Q}{\pi N} \right) ^{3/2}
\frac{S^{3/2}}{S^{3/2}+Q^{3/2}}.
\label{eq:Hayward-mu-SQN}
\end{align}
It is obvious that, 
under the rescaling $S\rightarrow \lambda S,~Q\rightarrow \lambda Q,~
N\rightarrow \lambda N$ of the additive variables $S,~Q,~N$, the mass $M$ 
scales as $M\rightarrow \lambda M$, whilst $T,~\Phi_{\rm H},~N$ remain unchanged. 
These homogeneity behaviors are expected for a standard extensive thermodynamic system.
Therefore, we can proceed using standard techniques for analyzing thermodynamic 
systems to study the behavior of the Hayward class BHs.

Let us look at the isocharge $T-S$ processes at fixed $N$. 
It is natural to seek for inflection points to see whether there could be 
any critical points. It turns out that the inflection point equations
\begin{align}
\left( \frac{\partial T}{\partial S} \right)_{Q,N}
=\left( \frac{\partial ^2T}{\partial S^2} \right) _{Q,N}=0
\end{align}
indeed have a solution with $N>0$ and $T>0$. We label the parameters 
$S,~Q,~T$ at the inflection point as $S_c,~Q_c,~T_c$, and their values are 
given as follows,
\begin{align}
Q_c &= \frac{\pi \alpha}{6L^2}
\frac{\left( A^{3/2}+1 \right) ^3}{A\left( 5A^{3/2}-1 \right)} N,
\label{Qc}\\
S_c &=A Q_c = \frac{\pi \alpha}{6L^2}
\frac{\left( A^{3/2}+1 \right) ^3}{\left( 5A^{3/2}-1 \right)} N,\\
T_c&=\frac{\sqrt{6}}{4\pi \alpha ^{1/2}}
\frac{4A^{3/2}-2}{\left( A^{3/2}+1 \right)^{3/2}\left( 5A^{3/2}-1 \right) ^{1/2}},
\end{align}
where
\begin{align}
A&=\left( \frac{8+3\sqrt{6}}{10} \right)^{2/3}.
\end{align}
Notice that $T_c$ is independent of $N$, which is a characteristic feature 
required in extensive thermodynamics. 

Using these ``critical'' values, we can introduce the relative 
electric charge, entropy and temperature as follows,
\begin{align}\label{eq:relative-parameters}
\mathcal{Q} =\frac{Q}{Q_c},\quad 
\mathcal{S} =\frac{S}{S_c},\quad 
\mathcal{T} =\frac{T}{T_c}
=\frac{\left( 5A^{3/2}-1 \right)}{4A^{3/2}-2}\frac{1}{\mathcal{S} ^{1/2}}
-\frac{\left( A^{3/2}+1 \right) ^3}{4A^{3/2}-2}
\frac{\mathcal{S}^{1/2}\mathcal{Q} ^3}{\left( A^{3/2}\mathcal{S}^{3/2}
+\mathcal{Q}^{3/2} \right) ^2}. 
\end{align}
Moreover, we also introduce the Helmholtz free energy $F=M-TS$ 
and its relative value $\displaystyle \mathcal{F}=\frac{F}{F_c}$. It follows that
\begin{align}
\mathcal{F}&=\frac{1}{B}\left[\frac{3A\left( 5A^{3/2}-1 \right)^{1/2}
\mathcal{S}^{1/2}}{\left( A^{3/2}+1 \right)^{3/2}}
+\frac{\mathcal{Q}^3\left( 5\mathcal{S}^{3/2}A^{3/2}
+2\mathcal{Q}^{3/2} \right) \left( A^{3/2}+1 \right)^{3/2}}
{\left(\mathcal{S}^{3/2}A^{3/2}+\mathcal{Q} ^{3/2} \right)^2 A^{1/2}
\left( 5A^{3/2}-1 \right)^{1/2}} \right],
\\
B&= \frac{20A^3+4A^{3/2}+2}{\left( A^{3/2}+1 \right)^{3/2} A^{1/2}
\left( 5A^{3/2}-1 \right) ^{1/2}}. 
\end{align}

\begin{figure}[!htb]
\centering
\begin{subfigure}[b]{0.4\textwidth}
    \includegraphics[width=\textwidth]{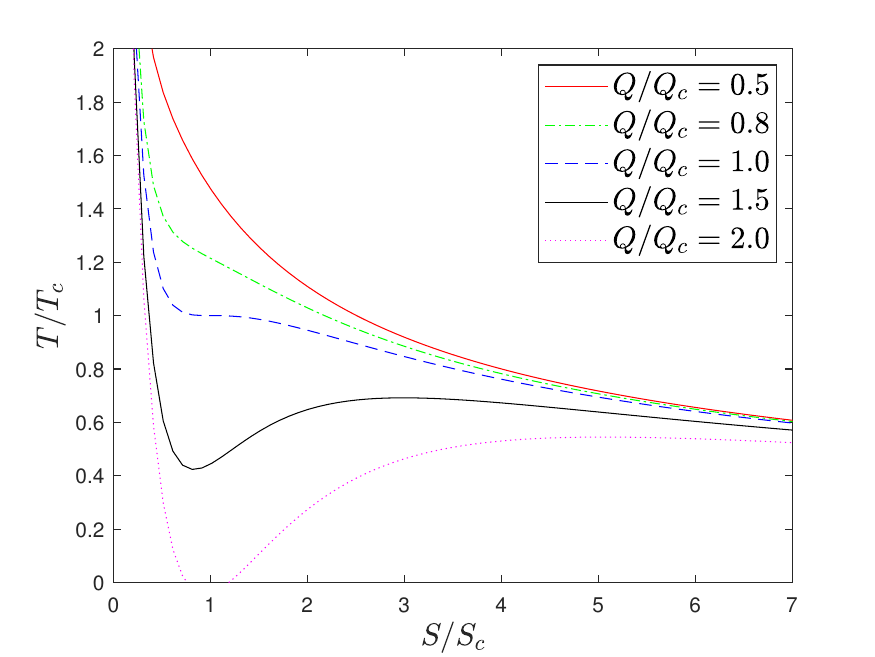}
   \caption{$\mathcal{T}-\mathcal{S}$ curve of Hayward BH}
   \label{fig:Hayward-T-S} 
\end{subfigure}
\begin{subfigure}[b]{0.4\textwidth}
   \includegraphics[width=\textwidth]{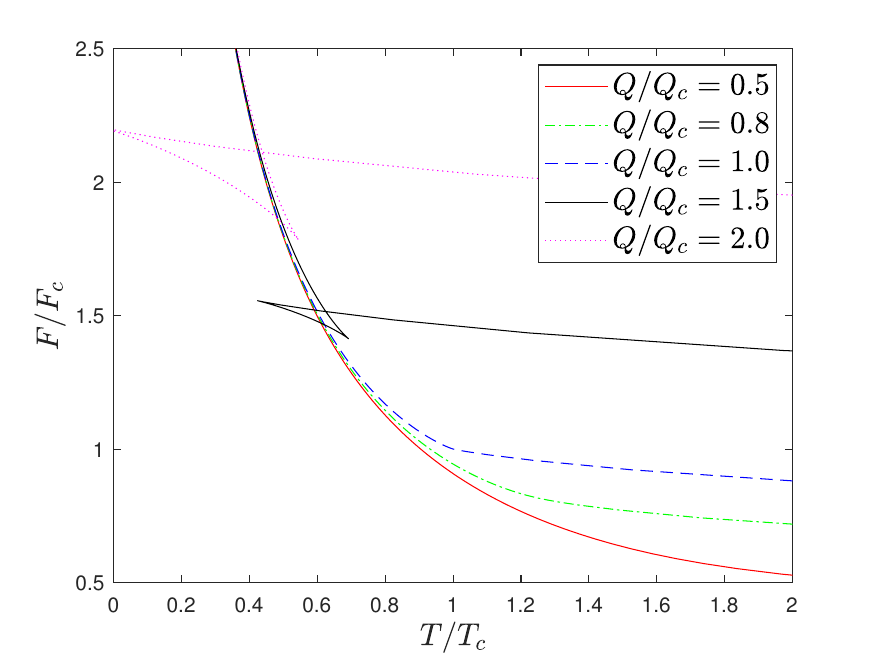}
   \caption{$\mathcal{F}-\mathcal{T}$ curve of Hayward BH}
   \label{fig:Hayward-F-T} 
\end{subfigure}
\caption{$\mathcal{T}-\mathcal{S}$ \eqref{fig:Hayward-T-S} and $\mathcal{F}-\mathcal{T}$ \eqref{fig:Hayward-F-T} curves of Hayward BH at fixed $Q,~N$.}
     
\end{figure}

Fig.\ \ref{fig:Hayward-T-S} and Fig.\ \ref{fig:Hayward-F-T} depict 
the $\mathcal{T}-\mathcal{S}$ and $\mathcal{F}-\mathcal{T}$ curves respectively
at fixed $Q_e,~N$. Contrary to the cases of RN-AdS and Kerr-AdS BHs which exhibit
two stable phases on the small and the large entropy ends, the 
$\mathcal{T}-\mathcal{S}$ and the $\mathcal{F}-\mathcal{T}$
curves for the Hayward class BHs seem to have been turned
upside-down. Consequently, there can be at most one stable phase 
for the Hayward class BHs which exists only for ``supercritical'' values of 
the electric charge in a limited range of the entropy (i.e. ranging from the local
minimum to the local maximum). The stability in this range of parameters 
can be justified by the positivity of the heat capacity 
\[
C_{Q,N}=\left(\frac{\partial M}{\partial T}\right)_{Q,N}
={T}\left[\left(\frac{\partial T}{\partial S}\right)_{Q,N}\right]^{-1}.
\] 
Therefore, strictly speaking, the inflection point described above 
is actually not a critical point for phase transitions but rather 
a critical point for the stable phase to appear. 

\section{RBHs as intermediate thermodynamic states}\label{se:Thought}

The analysis for the $\mathcal{T}-\mathcal{S}$ processes made in Section 
\ref{se:Thermodynamic} works for the complete Hayward class BHs 
involving both singular and regular states. However, where exactly the regular
state appears in the $\mathcal{T}-\mathcal{S}$ processes is not 
touched upon. The purpose of this section is to identify the position of 
the regular states in concrete $\mathcal{T}-\mathcal{S}$ processes. 

Before dwelling into details, let us first make some discussions about the 
regularity conditions and the number of roots for the shape function $f(r)$.
Let us recall that one of the regularity condition, 
\begin{align}
M_g= \frac{q^3}{G\alpha}, \label{reg-M-alpha-q}
\end{align}
is a constraint between the mass and the charge. Such a constraint resembles the 
saturated Bogomol'nyi bound 
\[
Q^2=4\pi \varepsilon_0 GM^2
\]
for the extremal RN spacetime. Such a similarity suggests a comparison 
between the Hayward class BHs and the RN BHs, which can be best illustrated by 
questing the number of roots for the shape function $f(r)$. 

Let us fix $s=3$ and rewrite the shape function \eqref{eq:general-f(r)-Hayward}
in terms of the rescaled parameters
$R=r/GM_g,~J=q/GM_g,~\tilde{\alpha}=\alpha/G^2M_g^2$ as
\begin{equation}\label{eq:evolution-metric}
f(R,J)=1-\frac{2\left( 1-\tilde{\alpha}^{-1}J^3 \right)}{R}
-\frac{2\tilde{\alpha}^{-1}J^3R^2}{R^3+J^3}.
\end{equation}
Meanwhile, the regularity condition \eqref{reg-M-alpha-q} can be re-expressed 
using the rescaled parameter as
\begin{align}
\tilde{\alpha}^{-1}J^3_{\rm reg}=1.
\label{5.3}
\end{align}
Therefore, inserting $J=J_{\rm reg}$ into eq.~\eqref{eq:evolution-metric}, 
we get the shape function for the regular Hayward BH,
\begin{equation}\label{eq:Hayward-regular-transform}
f(R,J_{\rm reg})=1-\frac{2R^2}{R^3+J^3_{\rm reg}}.
\end{equation}

\begin{figure}[!htb]
    \centering
    \includegraphics[width=0.4\linewidth]{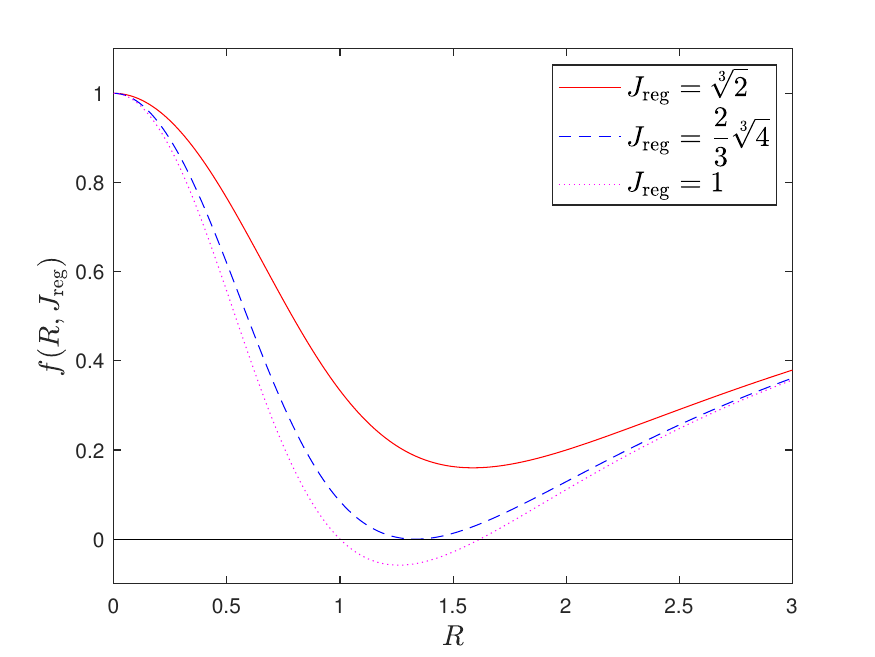}
    \caption{the curve of $f(R,J_{\rm reg})$ with different value of $J_{\rm reg}$.}
    \label{fig:Hayward-regular-f(r)}
\end{figure}

It follows that the number of roots for $f(R,J_{\rm reg})$ varies 
according to different choices of the parameter $J_{\rm reg}$. 
Fig.\ \ref{fig:Hayward-regular-f(r)} depicts the curves of $f(R,J_{\rm reg})$ 
versus $R$ for different choices of $J_{\rm reg}$. It turns out that
for $J_{\rm reg}>\dfrac{2}{3}\sqrt[3]{4}$, $f(R,J_{\rm reg})$ has no real root, 
implying that the spacetime contains no event horizon. 
When $J_{\rm reg}=\dfrac{2}{3}\sqrt[3]{4}$, $f(R,J_{\rm reg})$ has a single root. 
For $0<J_{\rm reg}<\dfrac{2}{3}\sqrt[3]{4}$, $f(R,J_{\rm reg})$ has two distinct 
real roots. Thus we see that in order that the shape function 
\eqref{eq:Hayward-regular-transform} describes a regular black hole, 
the parameter $J_{\rm reg}$ needs to have an upper bound 
which equals $\dfrac{2}{3}\sqrt[3]{4}$. 

It is meaningful to rewrite $J_{\rm reg}$ in terms of the relative electric 
charge $\mathcal{Q}$ defined in eq.~\eqref{eq:relative-parameters},
\begin{equation}\label{eq:J-Q_reg}
J_{\rm reg}=\frac{q}{GM_{\rm reg}}=\frac{\alpha}{q^2}
=\frac{6A\left( 5A^{3/2}-1 \right)}{\left( A^{3/2}+1 \right)^3\mathcal{Q}}.
\end{equation}
Using this equation, the upper bound for $J_{\rm reg}$ can be turned into 
the lower bound for $\mathcal{Q}$,
\begin{align}
\mathcal{Q}=\frac{6A\left( 5A^{3/2}-1 \right)}{J_{\rm reg} 
\left( A^{3/2}+1 \right) ^3 }\geq \frac{6A\left( 5A^{3/2}-1 \right)}
{\frac{2}{3}\sqrt[3]{4}\left( A^{3/2}+1 \right)^3}. 
\label{Qbound57}
\end{align}
Therefore, in order to have a regular black hole state, the spacetime needs 
to be sufficiently charged.

Now let us proceed to study the thermodynamic stability near the regular state. 
First of all, let us substitute eq.\eqref{rHq} together
with the regularity condition \eqref {reg-M-alpha-q} 
into the equation $f(r_H)=0$ 
and solve the result for $\alpha$. This gives
\begin{equation}\label{eq:requirement-alpha-Hayward}
\alpha =\frac{2q^3 r_{\rm H}^2}{r_{\rm H}^3+q^3}
=\frac{L^2}{\pi N}\frac{2Q^{3/2}S}{S^{3/2}+Q^{3/2}}
=\mathrm{const}.
\end{equation}
Inserting this result into eq.\eqref{eq:Hayward-T-SQN} we get 
\begin{align}
T_{\rm reg}&=\frac{N^{1/2}}{4\pi ^{1/2}L\,S^{1/2}}\frac{S^{3/2}
-2 Q^{3/2}}{S^{3/2}+Q^{3/2}} 
= \frac{N^{1/2}}{4\pi ^{1/2}L\, S^{1/2}}
\frac{X-2}{X+1},
\label{TregS}
\end{align}
where the dimensionless parameter $X$ is defined as
\[
X\equiv \frac{S^{3/2}}{Q^{3/2}}
=\left(\frac{\mathcal{S} A}{\mathcal{Q}}
\right)^{3/2}.
\]
The positivity of the temperature requires $X>2$. 

It is important to notice that the heat capacity of
the RBHs {\em should not} be calculated using 
eq.~\eqref{TregS}, because,  due to the constraint \eqref{eq:requirement-alpha-Hayward}, 
$S$ is not independent of $Q$ in the regular states. 
To calculate the heat capacity of the 
regular states, one should employ eq.~\eqref{eq:Hayward-T-SQN}
to calculate the partial derivative
$\displaystyle\left( \frac{\partial T}{\partial S} \right) _{Q,N}$, 
and then evaluate the result at the regular states. 
The final result reads
\begin{align}
\left. \left( \frac{\partial T}{\partial S} \right) _{Q,N}\right|_{\rm reg}
&=-\frac{N^{1/2}}{8\pi ^{1/2}L\,S^{3/2}}
\frac{S^3-13 S^{3/2}Q^{3/2}+4Q^3}
{\left( S^{3/2}+Q^{3/2} \right) ^2}\nonumber\\
&= -\frac{N^{1/2}}{8\pi ^{1/2}L\,S^{3/2}}
\frac{X^2-13X+4}{(X+1)^2}.
\end{align}
It is evident that the thermodynamic stability of the RBHs
requires $X^2-13X+4<0$. Combined with 
the requirement $X>2$, we find that the RBHs are 
thermodynamically stable only when 
\begin{align}
2<X<X_0 \equiv \dfrac{13+\sqrt{153}}{2}.
\label{StX}
\end{align}

Combining eqs.~\eqref{Qc}
with \eqref{eq:requirement-alpha-Hayward}, we get
\begin{align}\label{eq:S_reg-Q_reg}
\frac{\mathcal{S}\mathcal{Q}^{3/2}}
{\mathcal{S}^{3/2}A^{3/2}+\mathcal{Q}^{3/2}}
=\frac{3\left( 5A^{3/2}-1 \right)}
{\left( A^{3/2}+1 \right) ^3}.
\end{align}
This gives an implicit relationship between the relative
entropy and the relative charge at the regular states. 
It can be seen that, as long as the bound \eqref{Qbound57}
is not saturated, eq.\eqref{eq:requirement-alpha-Hayward}
regarded as an algebraic equation for $\mathcal{S}$ 
possesses two positive roots, among which only the bigger one 
corresponds to the black hole event horizon. 
Consequently, there is precisely one regular state 
on each isocharge $\mathcal{T}-\mathcal{S}$ curve.

For the bigger root of $\mathcal{S}$, it can be checked that 
$X$ is single-valued in $\mathcal{Q}$ and is 
monotonically decreasing as $\mathcal{Q}$ decreases. 
After some algebra, the thermodynamic stability condition 
\eqref{StX} can be turned into the following 
constraint condition over $\mathcal{Q}$,
\begin{align}
\dfrac{6A\left( 5A^{3/2}-1 \right)}
{\sqrt[3]{32/27}\left( A^{3/2}+1 \right)^3} \equiv \mathcal{Q}_{L}
<\mathcal{Q}<\mathcal{Q}_U
\equiv \dfrac{X_0+1}{X_{0}^{2/3}}
\dfrac{3A\left( 5A^{3/2}-1 \right)}
{\left( A^{3/2}+1 \right) ^3}.
\label{stabilitycond}
\end{align}
Both the upper and lower bounds for $\mathcal{Q}$ 
can be approximated by float point numbers, i.e.
\begin{align}
\mathcal{Q}_L\simeq 3.09133, \quad \mathcal{Q}_U \simeq 4.11553.
\end{align}

\begin{figure}[!h]
\centering
\begin{subfigure}[b]{0.42\textwidth}
    \includegraphics[width=\textwidth]{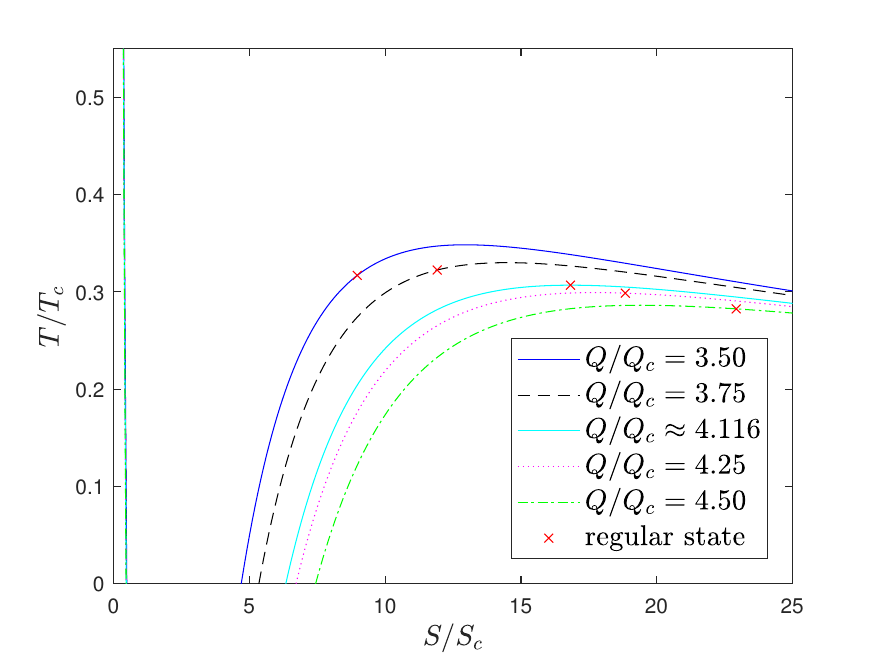}
   \caption{$\mathcal{T}-\mathcal{S}$ curves with regular states marked}
   \label{fig:Hayward-T-S2} 
\end{subfigure}
\hspace{2em}
\begin{subfigure}[b]{0.42\textwidth}
    \centering
    \includegraphics[width=\textwidth]{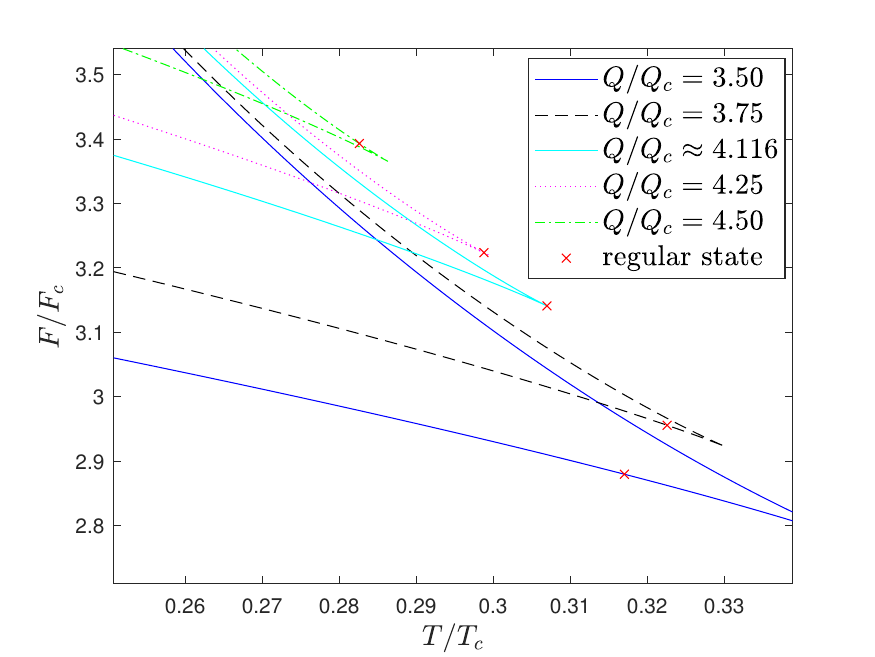}
    \caption{$\mathcal{F}-\mathcal{T}$ curves with regular states marked}
    \label{fig:Hayward-F-T-sd2}
\end{subfigure}
\caption{$\mathcal{T}-\mathcal{S}$ \eqref{fig:Hayward-T-S2} 
and $\mathcal{F}-\mathcal{T}$ 
\eqref{fig:Hayward-F-T-sd2} curves of Hayward BHs with regular states marked. 
The curves are intentionally zoomed-in in order to have a better resolution
to show the exact position of the regular states on the curves.}
\label{fig:curve-regular-state}
\end{figure}

\begin{figure}
    \centering
    \includegraphics[width=0.4\textwidth]{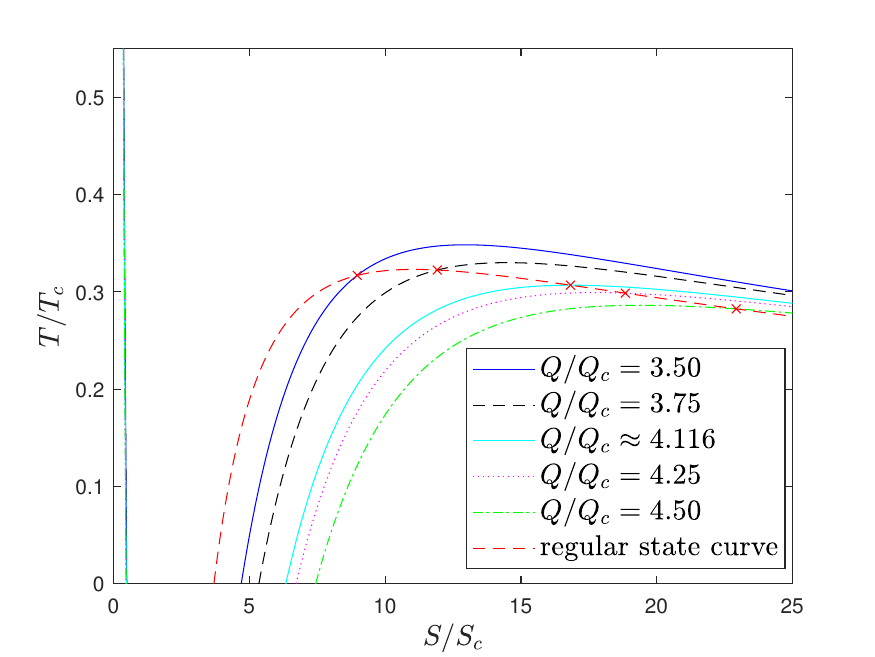}
    \caption{$\mathcal{T}-\mathcal{S}$ process between regular states cannot be isocharge process.}
    \label{fig:Hayward-T-S-regular-curve}
\end{figure}

Figs.~\ref{fig:Hayward-T-S2} and \ref{fig:Hayward-F-T-sd2}
depict the $\mathcal{T}-\mathcal{S}$ and 
$\mathcal{F}-\mathcal{T}$ curves for certain choices of 
$\mathcal{Q}$ with the regular states marked by a cross
($\times$). The selected values of $\mathcal{Q}$ are such that 
some of the values lie in between $\mathcal{Q}_L$ and $\mathcal{Q}_U$ 
and some exceed $\mathcal{Q}_U$. 
It can be seen that for those $\mathcal{Q}$ that lie 
in between the bounds, the corresponding regular states 
appear on the increasing segment of the 
$\mathcal{T}-\mathcal{S}$ curve, signifying thermodynamic stability.
On the other hand, for those $\mathcal{Q}$ that exceed 
the upper bound $\mathcal{Q}_U$, the corresponding regular states 
appear on the decreasing segment of the 
$\mathcal{T}-\mathcal{S}$ curve, signifying 
thermodynamic instability. Similar conclusion can also 
be drown from the $\mathcal{F}-\mathcal{T}$ curves, on which 
stable regular states appear on the lowest branch of 
$\mathcal{F}$, while the unstable ones do not. 

Due to the non-independence between $\mathcal{S}$ 
and $\mathcal{Q}$ for the regular states, any 
$\mathcal{T}-\mathcal{S}$ process
leading from one regular state to another cannot be 
an isocharge process. This is illustrated in 
Fig.~\ref{fig:Hayward-T-S-regular-curve}.

Before closing this section, let us remark that,
according to eqs.~\eqref{5.3} and \eqref{eq:J-Q_reg}, we have, 
at the regular states, 
\begin{align}\label{eq:alpha-Q_reg}
\tilde{\alpha}=J^3_{\rm reg}
=\left( \frac{6A\left( 5A^{3/2}-1 \right)}
{\left( A^{3/2}+1 \right)^3\mathcal{Q}} \right) ^3.
\end{align}
The thermodynamic stability condition for the regular states 
can be reformulated by use of the parameter $\tilde{\alpha}$, i.e.
\begin{align}
&\tilde \alpha_L < \tilde\alpha <\tilde\alpha_U,
\label{boundalpha}
\end{align}
wherein 
\begin{align}
&\tilde \alpha_L = \dfrac{16(13+\sqrt{153})^2}{(15+\sqrt{153})^3}
\approx 0.50228,\quad \alpha_U = \frac{32}{27}\approx 1.18519.
\end{align}
The bound \eqref{boundalpha} follows from the 
substituting eq.\eqref{eq:alpha-Q_reg} into
\eqref{stabilitycond}. For $\tilde{\alpha}$ in between the above bounds, 
the corresponding regular state will be thermodynamically stable.

\section{Quasinormal frequencies and dynamic stability}
\label{sec:qnf}

Besides thermodynamic stability, the dynamic stability of BHs is also 
important in order to have a full understanding about the behaviors 
of BHs under perturbation. The standard way for analyzing dynamic stability 
of BHs is to calculate the quasinormal frequencies (QNFs) of the BHs 
under small perturbation. Such analysis has been done for some of the 
known RBHs in Refs.~\cite{Bronnikov:2012ch,Konoplya:2022hll,
Konoplya:2023ahd,Konoplya:2023ppx}. 

In this section, we will use the 13th-order WKB method under 
the Pad\'e approximation \cite{Matyjasek:2017psv,Konoplya:2019hlu} to 
calculate the quasinormal frequencies (QNFs) of the electrically charged 
Hayward class BHs under the perturbation of a massless scalar field. 
We assume that the only parameter which changes as the BH evolves 
is $J$, which, for fixed black hole mass, corresponds to the change in charge. 
What we actually do is to calculate the QNFs for the whole class of 
BHs with the shape function \eqref{eq:evolution-metric} at fixed values of 
$\tilde\alpha$ without imposing the regularity condition \eqref{5.3}, and then 
look at the results at the particular choices of $J$ which obey the 
regularity condition. This makes a difference from previous works 
\cite{Bronnikov:2012ch,Konoplya:2022hll,Konoplya:2023ahd,Konoplya:2023ppx},
because those works considered exclusively the QNFs {\em within 
the regular BH configurations}. 

The massless Klein-Gordon equation reads
\begin{equation}
\frac{1}{\sqrt{-g}} \partial_\mu \sqrt{-g} g^{\mu \nu} \partial_\nu \Psi=0 .
\end{equation}
Since the spacetime is spherically symmetric, it is natural to make a separation 
of variables which takes the spherical harmonic functions 
$Y_{lm}(\theta, \varphi)$ as a factor,
\begin{equation}
\Psi(t,r,\theta,\varphi)
=\re^{-\mi \omega_{l} t} Y_{lm}(\theta, \varphi) \frac{\phi_l(r)}{r}.
\end{equation}
Then the Klein-Gordon equation can be reduced into the Sch\"odinger-like equation
for the radial factor $\phi_l(r)$, which, for brevity, will be written simply as 
$\phi(r)$:
\begin{equation}
\label{eq:SE-like}
\frac{\rd ^2 \phi}{\rd  r_*^2}+\omega^2 \phi=V_{\text {eff}}(r) \phi,
\end{equation}
where the Regge-Wheeler coordinate $r_*$ is defined via 
$\dfrac{\dif r_{*}}{\dif r}=\dfrac{1}{f(r)}$, and the effective potential
takes the form
\begin{equation}
V_{\mathrm{eff}}(r)=f(r)\left[\frac{l(l+1)}{r^2}+\frac{f^{\prime}(r)}{r}\right].
\end{equation}

The quasinormal modes are defined by the solution of eq.~\eqref{eq:SE-like} 
obeying the boundary conditions
\begin{equation}
\phi(r) \sim \re^{ \pm \mi \omega r_*}, \quad r_* \rightarrow \pm \infty.
\end{equation}
In previous models studies on RBHs, there is also an inherent correlation 
between the thermodynamics and dynamics \cite{Miao:2020uft,Li:2021qim}. 
Some of studies indicates that the thermodynamic stability and dynamic stability 
can be different for the same BH solution with the same parameter set.
We would like to check whether such phenomena also occurs for the regular states. 
For this purpose, we take two different values for the parameter $\tilde\alpha$, 
i.e. $\tilde\alpha =1$ in the thermodynamically stable range and $\tilde\alpha=0.1$
in the thermodynamically unstable range, and study the corresponding 
dynamic stability.

\begin{figure}[!htb]
     \centering
     \begin{subfigure}[b]{0.31\textwidth}
         \centering
         \includegraphics[width=\textwidth]{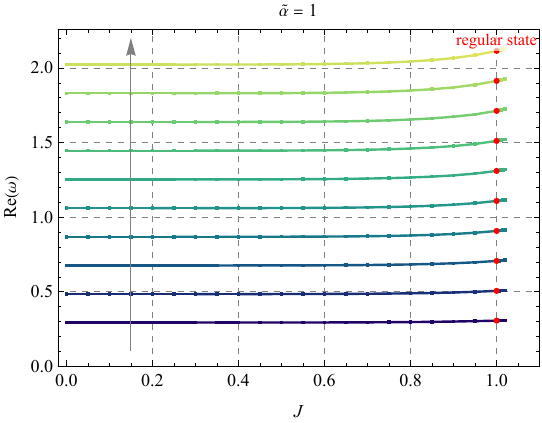}
         \caption{$\Re(\omega)$ versus $J$}
         \label{fig:qnf-re-J}
     \end{subfigure}
     \begin{subfigure}[b]{0.31\textwidth}
         \centering
         \includegraphics[width=\textwidth]{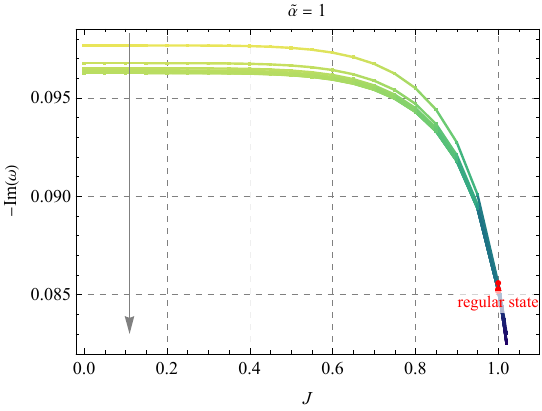}
         \caption{$-\Im(\omega)$ versus $J$}
         \label{fig:qnf-im-J}
     \end{subfigure}
\begin{subfigure}[b]{0.31\textwidth}
     \centering
    \includegraphics[width=\textwidth]{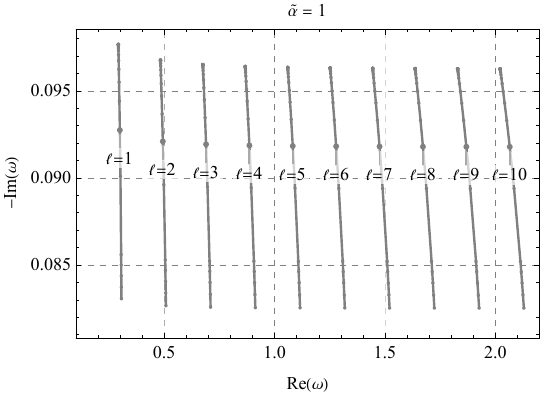}
       \caption{QNFs in the complex plane}
        \label{fig:qnf-l}
\end{subfigure}
\captionsetup{width=.9\textwidth}
\caption{QNFs at different values of $l$ with $\tilde\alpha=1$. 
The arrows in Panel \ref{fig:qnf-re-J} and \ref{fig:qnf-im-J} 
point to the increasing direction of $l$, which ranges from $1$ to $10$. }
\label{fig:qnf}
\end{figure}

It should be reminded that for fixed $\tilde\alpha$, the 
event horizon equation $f(R,J)=0$ implies that there is an upper bound for $J$. 
Therefore, any $\omega-J$ curve should have 
an end point at $J=J_{\rm max}$.

Fig.~\ref{fig:qnf-re-J} and Fig.~\ref{fig:qnf-im-J} respectively 
depict the real and imaginary parts of the QNFs as functions in $J$ 
at different values of $l$, with the regular states marked by round dots. 
It can be seen that the imaginary parts of the QNFs at 
the regular states are always  negative, signifying dynamic stability. 
Meanwhile, away from the regular states, the real part of the QNFs
increases with $J$ at fixed $l$, and also with $l$ at fixed $J$, 
while the absolute value of the negative imaginary part decreases 
with $J$ at fixed $l$, and also with $l$ at fixed $J$. 

Fig.~\ref{fig:qnf-l} depicts the QNFs on the complex plane with the imaginary
axis turned upside down. One can see that for sufficiently large $l$, the upper and lower 
bounds for the imaginary parts gradually become constant. 

\begin{figure}[!htb]
     \centering
     \begin{subfigure}[b]{0.31\textwidth}
         \centering
         \includegraphics[width=\textwidth]{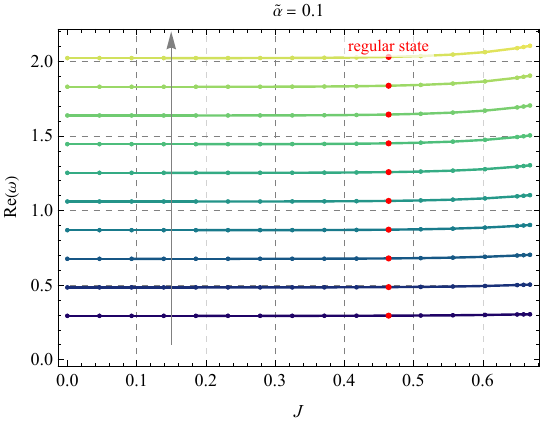}
         \caption{$\Re(\omega)$ versus $J$}
         \label{fig:qnf-re-J-p}
     \end{subfigure}
     \begin{subfigure}[b]{0.31\textwidth}
         \centering
         \includegraphics[width=\textwidth]{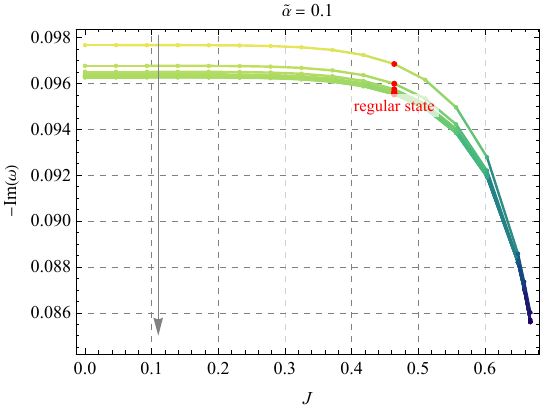}
         \caption{$-\Im(\omega)$ versus $J$}
         \label{fig:qnf-im-J-p}
     \end{subfigure}
     \begin{subfigure}[b]{0.31\textwidth}
     \centering
    \includegraphics[width=\textwidth]{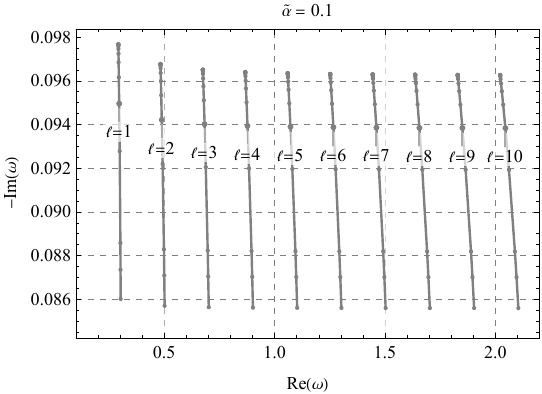}
       \caption{QNFs in the complex plane}
        \label{fig:qnf-l-p}
\end{subfigure}
      \captionsetup{width=.9\textwidth}
\caption{QNFs at different values of $l$ with $\tilde\alpha=0.1$. 
The arrows in Panel \ref{fig:qnf-re-J-p} and \ref{fig:qnf-im-J-p} 
point to the increasing direction of $l$, which ranges from $1$ to $10$. }
\label{fig:qnf-p}
        
\end{figure}

One might expect that something different might happen when $\tilde\alpha$ takes 
values beyond the scope \eqref{boundalpha}. However, the fact is not the case, 
as is shown in Fig.~\ref{fig:qnf-p}, which are 
parallel to Fig.~\ref{fig:qnf}, 
but with $\tilde\alpha =0.1$. The negativity of the imaginary parts 
of the QNFs as depicted in Fig.~\ref{fig:qnf-im-J-p} indicates that, 
the regular states in this case, although being thermodynamically unstable, 
are actually dynamically stable under massless scalar perturbations.
Besides, all qualitative behaviors of the QNFs with $\tilde\alpha=0.1$ 
are basically identical to the case with $\tilde\alpha=1$. Therefore, our work indicates
once again that, for the same BH solution with the same parameter sets, the thermodynamic
stability and the dynamic stability need not to be identical.

\section{Conclusion and outlook}
\label{se:Conclusion}

The RPS formalism in which the Euler relation is put at the center position
appears to hold for RBHs. However, the requirement of parameter independence 
in BH thermodynamics calls for the view of treating RBHs as intermediate 
thermodynamic states in a larger class of BHS involving both regular and singular 
states. This idea is verified in several classes of BHs incolving regular states, including 
the Bardeen(-AdS) class, Hayward class and a class of BH solutions 
of the model introduced in \cite{li2023regular}. The idea that 
RBHs should be viewed as intermediate states has not been considered seriously before, 
and thus our approach is totally different from the previous researches such as 
\cite{Zhang:2016ilt,Fan:2016hvf,Fan:2016rih} on the thermodynamics of RBHs. 

Following our proposal, it is shown that, for the electrically charged 
Hayward class BHs, the regular states may be thermodynamically either stable or 
unstable, depending on the amount of charges carried by the BHs. 
However, by studying the QNFs of the whole Hayward class BHs, it is shown that,
even for the thermodynamically unstable regular states, the dynamic stability 
can still hold, at least under massless scalar perturbations. 
This gives further evidence to the view that the thermodynamic and dynamical 
stabilities for the same BH solution with the same parameter set need not to
be identical. 

Last but not the least, the successful application of the RPS formalism to the several 
classes of BHs involving regular states indicates the strength and wide 
applicability of this formalism itself, which seems to be worth of further 
explorations.

\section*{Acknowledgement}
This work is supported by the National Natural Science Foundation 
of China under Grant No. 12275138.

\section*{Data Availability Statement} 

This research makes no use of new data. The QNFs analyzed in Section \ref{sec:qnf}
is completely based on numerical solutions to eq.\eqref{eq:SE-like}, 
for which the procedure is standard. 

\section*{Declaration of competing interest}

The authors declare no competing interest.



\providecommand{\href}[2]{#2}\begingroup\raggedright\endgroup

\end{document}